\documentclass[graybox]{svmult}
\usepackage{wrapfig}
\usepackage{ mathrsfs }

\usepackage{sidecap}
\usepackage{amsmath}
\usepackage{calc}
\usepackage{mathptmx}       
\usepackage{helvet}         
\usepackage{courier}        
\usepackage{type1cm}        
\usepackage{relsize}
 
\usepackage{ wasysym }
\usepackage{makeidx}         
\usepackage[pdftex]{graphicx}        
\usepackage{multicol}        
\usepackage[bottom]{footmisc}

\makeindex             

\begin{document}

\title*{Gravitational Radiation from Compact Binary Pulsars}
\author{John Antoniadis}
\institute{Max-Planck-Institut f\"{u}r Radioastronomie \at Auf dem H\"{u}gel 69, 53121, Bonn, Germany, \email{jantoniadis@mpifr-bonn.mpg.de}}
%
%
\maketitle

\abstract*{An outstanding question in modern Physics is whether general relativity (GR) is a complete description of gravity among bodies at macroscopic scales. Currently, the best experiments supporting this hypothesis are based on high-precision timing of radio pulsars. 
This chapter reviews recent advances in the field with a focus on compact binary millisecond pulsars with white-dwarf (WD) companions. These systems -- if modeled properly -- provide an unparalleled test ground for physically motivated alternatives to GR that deviate significantly in the strong-field regime. Recent improvements in observational techniques and advances in our understanding of WD interiors have enabled  a series of precise mass measurements in such systems. These masses, combined with high-precision radio timing of the pulsars, result to stringent constraints on the radiative properties of gravity, qualitatively very different from what was available in the past. }

\abstract{An outstanding question in modern Physics is whether general relativity (GR) is a complete description of gravity among bodies at macroscopic scales. Currently, the best 
experiments supporting this hypothesis are based on high-precision timing of radio pulsars. 
This chapter reviews recent advances in the field with a focus on compact binary millisecond pulsars with white-dwarf (WD) companions. These systems -- if modeled properly -- 
provide an unparalleled test ground for physically motivated alternatives to GR that deviate significantly in the strong-field regime. Recent improvements in observational techniques 
and advances in our understanding of WD interiors have enabled  a series of precise mass measurements in such systems. These masses, combined with high-precision radio timing of 
the pulsars, result to stringent constraints on the radiative properties of gravity, qualitatively very different from what was available in the past. }

\section{Introduction}
\label{intro}
The centennial of Einstein's General Theory of Relativity (GR) will also mark the beginning of a new era in experimental gravity. Ground-based antennas such as advanced LIGO and 
VIRGO will be put in use to observe mergers of compact objects through their gravitational-wave (GW) emission. This new endeavor builds on the remarkable success  of 
GR in describing gravitational phenomena and its wide-spread use in contemporary astrophysics. Following a hibernation period after Einstein's original publication \cite{gr1,gr1916}, 
discoveries such 
as quasars, {X}-ray binaries and neutron stars (NSs), as well as the emergence of  observational cosmology made GR a vital part of astronomers' toolbox. At the same time however, it 
was realized that GR -- a non-renormalizable field theory -- was difficult to reconcile with the Standard Model (SM) of Particle Physics \cite{feynmangr}. Furthermore, some key 
features of modern cosmology like dark matter and dark energy forced scientists to consider the possibility that the theory is only valid for a limited range of energy scales. This 
motivated the investigation of numerous alternative theories, inevitably leading  to the need for high-precision tests of gravity \cite{will94,will06}.  

During the past 60 years GR has been tested in numerous different settings and emerged victorious in every single case. An important chapter in precision tests  opened with 
the discovery of the first binary pulsar by Russell Hulse and Joe Taylor in 1974 \cite{ht75a}. For the first time, this system allowed to test the self consistency of GR for orbiting strongly 
self-gravitating objects. Most importantly, it led to the first experimental proof of gravitational radiation through the detection of the system's orbital decay 
\cite{tfm79,wt81,tw82,wt84,wnt10}. 
\begin{svgraybox}
Precision tests of gravity in the Solar System include: 
\begin{itemize}
\item The perihelion-advance of Mercury \cite{gr0,vfl+14}, 
\item Light deflection in Sun's and Jupiter's gravitational fields \cite{eclipse,ll11} ,
\item Strong Equivalence Principle (SEP) tests and the de Sitter precession of the Moon's orbit \cite{lrr10,hmb10},
\item Lens-Thirring effect in satellite orbits \cite{lteffect},
\item Relativistic spin precession of gyroscopes  \cite{GravityProbeB},
\item Shapiro delay due to the Sun's gravitational field \cite{shapiro,bit03}.
\end{itemize}
Most of these tests have now reached a precision of $10^{-4}$ or better  (see \cite{will94,will14}).
\end{svgraybox}
Presently this effect has been measured in ten more binary pulsars and a  double white-dwarf (WD) system (Table\,1). These systems allow for a range of strong-field and radiative tests of  gravity in a  regime where many alternatives predict significant deviations from the GR predictions. This chapter summarizes recent advances in the field focusing on compact binary pulsars with WD companions which, in the context of many alternative theories of gravity, should radiate significant amounts of dipolar GWs. 
 This chapter is a brief review of a few recent studies from the point of view of an experimentalist. The text is split in two major parts, background theory (Sections\,2--6) and results (Sections\,7--9). For  thorough reviews on theoretical aspects the reader is referred to \cite{will14,wex}.

\section{Pulsars and Pulsar Timing}  
\label{sec:2}
Pulsars play an important role in precision tests of gravity for two reasons: Firstly, they contain extremely strong gravity fields (surface strengths of order $GM/c^2R \simeq 0.2$, compared to $10^{-6}$ for the Sun). Secondly, they possess clock-like properties that enable the detection of tiny orbital effects with high precision.

The regularity of spin periods  for some of the known pulsars rivals the best atomic clocks on Earth. However, radio pulses \emph{do not} arrive at a constant rate at our telescopes and this is exactly what allows to measure most of their properties. Figure\,1 displays the measured $P_{\rm spin}$ of all known 2200 pulsars in the Galaxy \cite{mhth05} against the most common change in the spin period -- the first spin-period derivative $\dot{P}_{\rm spin}$  \cite{lk05}. Pulsars tend to cluster in two distinct regions of the $P_{\rm spin}-\dot{P}_{\rm spin}$ diagram.  
\begin{figure}[h]
  \begin{center}
   \label{fig:1}
     \caption{A $P_{\rm spin}-\dot{P}_{\rm spin}$ diagram of the known radio pulsars in the Galactic disk (adapted from \cite{ewan}). The secular change of the spin period is caused by energy   emission from the star's rotating electromagnetic field. Dotted lines depict representative values for the spin-down energy and the magnetic field strength.  The blue line corresponds to a critical value of the spin-down energy bellow which electromagnetic emission is switched off. For more details see \cite{lk05}.}
    \includegraphics[width=0.77\textwidth]{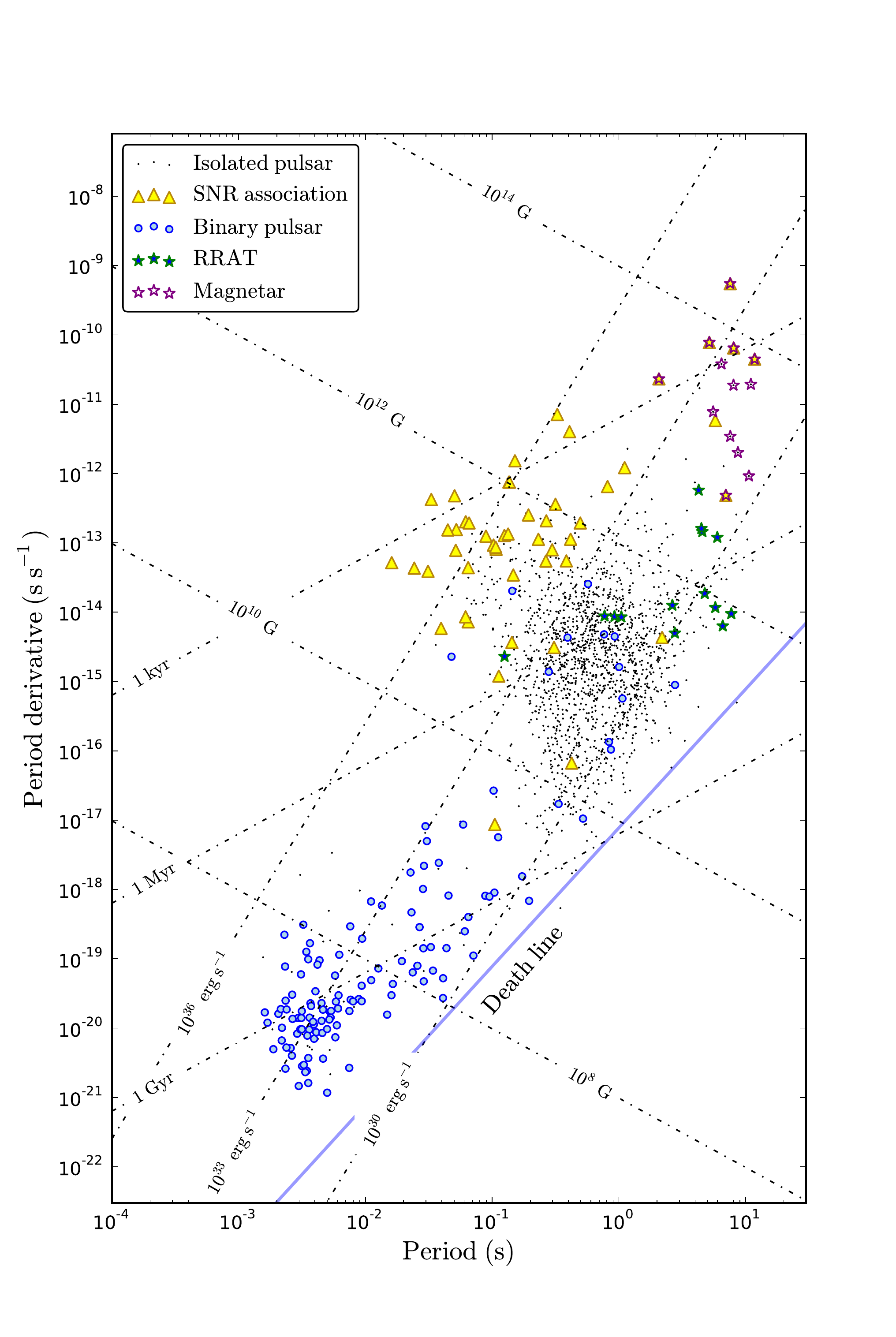}
  \end{center}
\end{figure}
There are ``normal'' pulsars with $(P_{\rm spin},\dot{P}_{\rm spin})\sim (1$\,s, 10$^{-15}$\,s\,s$^{-1}$) and ``millisecond'' pulsars  with $(P_{\rm spin},  \dot{P}_{\rm spin})   \sim (3$\,ms, $10^{-20}$\,s\,s$^{-1})$. The latter are usually much more stable, i.e. there is no other intrinsic rotation change other than the secular variation described above. Conveniently, millisecond pulsars are also more likely to be
found in a binary system, a consequence of their evolution, which is
influenced by accretion of mass from a binary companion. Most  systems  discussed here belong to this second category. 
Because of their intrinsic stability, millisecond pulsars can be used as experimental clocks   to infer the properties of  spacetime around them. 
The technique used to achieve this is called pulsar timing and is briefly described below.  

\subsection{Pulsar Timing}
For most astronomical objects, the relative motion of the source along the line of sight is determined from Doppler measurements of the emitted signals. For
pulsars however, one can do much better. Here, the signal of interest is the beamed electromagnetic radiation sweeping the telescope once per rotation.
The pulse profile –-- especially for millisecond pulsars –-- occupies a very small fraction of the rotational phase, typically few percent. This allows to determine a precise pulse
time of arrival (TOA) relative to a fiducial point on the profile (e.g. its peak), which
corresponds to a constant physical latitude on the NS. Thus, since the time difference between any pair of TOAs corresponds to an
integer number of
rotations, the parametric model describing deviations from constant spin can be optimized to take into
account all pulsar rotations between two measurements. This phase-coherent approach of pulsar timing enables inference of
rotational and orbital parameters orders of magnitude more precise than traditional Doppler techniques.

The timing model that relates observed TOAs with proper emission times has to  account  for the accelerating reference frame of the observer, propagation delays caused by the interstellar medium and the motion of the pulsar and the radio telescope relative to the Solar System Barycentre (SSB):
\begin{equation}
 t_{0} = t_{\rm telesope}  +D/f^2 - \Delta _{\rm R} -\Delta_{\rm E} - \Delta_{\rm S} + \Delta_{\rm PSR}
\end{equation}
Here, the first term accounts for frequency-dependent delays of the signal caused by the interstellar medium. The first  three $\Delta$ terms account for the Roemer, Einstein and Shapiro delays in the Solar System and transform the TOA at the non-inertial terrestrial reference frame of the observer  to a TOA at the inertial SSB frame. The last term of the equation encodes information about the pulsar's  orbital motion  and is what gives access to most gravity tests. 

\subsection{Binary Pulsars}

For binary pulsars, the time delay due to the orbital motion, $\Delta_{\rm PSR}$, can be modeled by means of the Keplerian parameters  and a set of post-Keplerian (pK) terms that  account for any possible deviation from the Newtonian orbit \cite{dt91,lk05}.  
Typically, these pK parameters include secular and higher-order variations of the Keplerian parameters ($\dot{e}$, $\dot{\omega}$, $\dot{x}$ etc.), two ``Shapiro delay'' terms, $r$ and $s$ that model variations of the light-travel time due to the curvature of spacetime around the binary companion \cite{shapiro,bt76,dd86,fw10b} and an ``Einstein delay'' term $\gamma$,  that accounts for time dilation effects due to the motion of the pulsar and the varying gravitational redshift along the orbit.  
For  boost-invariant gravity theories, in the slow motion approximation, the pK parameters  become functions of the  orbital elements, the inertial  masses of the system and the equation-of-state (EoS) describing the properties of the stellar interior  \cite{dd85,dd86}. 

In GR, even though the  spacetime curvature can be extremely large, the field equations satisfy exactly the strong equivalence principle (SEP) leading to an effacement of the internal stellar structure. In other words, internal coupling terms (e.g. binding energy)  can be absorbed in redefinitions of the masses, so that the pK parameters are only functions of the  masses and orbital elements. In the first post-Newtonian approximation level $\mathscr{O}(v^2/c^2)$, one finds the following  deviations from the Keplerian orbit:

\begin{equation}
\dot{\omega} = 3 \left( \frac{P_{\rm b}}{2\pi} \right)^{-5/3} \left( T_{\odot}M \right)^{2/3} \left(1-e^2 \right)^{-1},
\end{equation}
\begin{equation}
r=T_{\odot}m_{\rm c},\,s=\sin i = x \left(\frac{P_{\rm b}}{2\pi}\right)^{-2/3} T_{\odot}^{-1/3} M^{2/3}m^{-1}_{\rm c}
\end{equation}
\begin{equation}
\gamma = e \left( \frac{P_{\rm b}}{2\pi} \right)^{1/3} T^{2/3}_{\odot} M^{-4/3} 
m_{\rm c} \left(m_{\rm p}+2m_{\rm c} \right),
\end{equation}
where $T_{\odot} \equiv G M_{\odot} /
c^3 = 4.925490947 \mu$s is a solar mass in time units and $M=m_1+m_2$ is the total mass of 
the binary. We note that the former set of equations refers to the intrinsic 
pK effects that can be extracted from the measured values after 
taking into account kinematic corrections \cite{dt91}.

Equations\,2 -- 4 relate observable, model-independent quantities with the a priori unknown masses. Whenever two pK parameters become measurable the masses are uniquely determined, of course withing the framework of a particular gravity theory. Thus,  the measurement of additional pK parameters become tests of the gravity theory, since their values are fixed from the already determined masses (see Figure\,2 for a graphical example in the case of the double pulsar).

\subsection{Gravitational Wave Emission}
Gravitational wave emission from accelerating masses is expected for theories in which gravity propagates with a finite speed. As we shall see below, detection and characterization of GWs is of fundamental importance for gravity tests as the radiative properties are a sensitive probe of the underlying gravity theory.  In GR, the loss of energy to GWs first enters the binary motion as a quadrupole term at the 2.5 post-Newtonian level [$\mathscr{O}(v^5/c^5)$]. The associated orbital-period decay, averaged over one orbit reads:

\begin{equation}
\dot{P}_{\rm b} = -\frac{192\pi}{5}  \left( 1 + \frac{73}{24}e^2 + \frac{37}{96}e^4 \right) \left(1-e^2 \right)^{-7/2} \left(\frac{2\pi \mathcal{M}}{P_{\rm b}}\right)^{5/3}T_{\odot}^{5/3},
\end{equation}
where $\mathcal{M} = (m_{\rm p} m_{\rm c})^{3/5} (m_{\rm p} + m_{\rm c})^{-1/5}$ is the so-called chirp mass of the system.

\begin{svgraybox}
In addition to the emission of GWs, the observed orbital decay of a binary pulsar can  be affected by several ``classical'' terms. 
These effects are either apparent (kinematic) or intrinsic to the system. \\
Kinematic effects include a  secular change due to the Galactic gravitational potential and the proper motion of the pulsar, $\mu$, on the sky:

\begin{equation}
\delta \dot{P}_{\rm b}^{\rm kin} = \frac{P_{\rm b}}{c}\left[ \mathbf{\hat{K}}_{0} \cdot \left( \mathbf{g}_{\rm PSR} - \mathbf{g}_{\odot} \right) + \mu d^2  \right],
\end{equation}
where $\mathbf{\hat K}_{0}$ is the unit vector pointing from the Earth to the pulsar and $\mathbf{g}_{\rm PSR,\odot},$ are the pulsar and Earth acceleration vectors in the Galactic gravitational potential.\\
Intrinsic orbital-period variations due to ``classical'' effects may arise from:
\begin{enumerate}
\item Changes in the companion star's quadrupole moment \cite{lvt+11,akh+13} that become apparent due to spin-orbit couplings,  
\item mass loss either from the pulsar or the companion, $\dot{P}_{\rm b}^{\dot{M}} =2\dot{M}/M$, where $M = m_{\rm p} + m_{\rm c}$ \cite{afw+13} and
\item an orbital decay caused by tides: 
\begin{equation}
\dot{P}_{\rm b}^{\rm T} = \frac{3 k \Omega_{\rm c}}{2\pi q (q + 1)} \left( \frac{R_{\rm c}P_{\rm b}\sin i}{x c} \right) ^2 \frac{1}{\tau_{\rm s}},
\end{equation}
derived from conservation of angular momentum. Here $\Omega_{\rm c}$ is the angular velocity of the companion, $\tau_{\rm s} = \dot{\Omega_{\rm c}}/\Omega_{\rm c}$ is the synchronization timescale and 
$k$ is related to the companion's moment of inertia, $k \equiv I_{\rm c}/(m_{\rm c}R_{\rm c}^2)$. $R_{\rm c}$ is the companion's radius and $q$ the system's mass ratio.
\end{enumerate}
For the systems discussed here these terms can either be taken into account or are much bellow the measurement uncertainties.

\end{svgraybox}

From Eq.\,6 we see that the back-reaction to the orbit depends sensitively on the orbital period and less so on the eccentricity and the stellar masses.
As of today, $\dot{P}_{\rm b}^{\rm intrinsic}$ has been detected in 11 binary pulsars and a double-WD system (Table\,1). All of these binaries have orbital periods $\le 1$\,day 
and span a broad range of masses and eccentricities. While all  systems are interesting in their own regard, only few  stand out as laboratories for strong gravity. The remaining of this text focuses on those, following a somewhat historical order.

\begin{table}
\resizebox{0.86\textwidth}{!}{\begin{minipage}{\textwidth}
\begin{center}
\label{tab:1}

\begin{tabular}{l|c|c|c|c|c|c|l|l|c|c}

& Comp. &$P_{\rm spin}$ & $P_{\rm b}$& $m_{\rm p}$ & $m_{\rm c}$ &  &$\dot{P}_{\rm b}^{\rm GR}$& $\dot{P}_{\rm b}^{\rm int}$& $d$ &\\ 
Name&Type &[ms] & [h] & [M$_{\odot}$] &  [M$_{\odot}$] & $e$&[$\times 10^{-12}$] & [$\times 10^{-12}$] &  [kpc] &Ref. \\

\hline 
J0737$-$3039 & PSR & 22.7 & 2.5 & 1.3381(7) & 1.2489(7) & 0.08  &$-1.24787(13)$  & $-1.252(17)$  & $1.15(22)$ &\cite{ksm+06,dbt09} \\ 
 
B1534+12 & NS & 37.9 & 10.1 & 1.3330(4) & 1.3455(4) & 0.27 &$-0.1366(3)$ & $-0.19244(5)$ & $0.7$ &\cite{sac+98,fst14}\\ 

J1756$-$2251 & NS & 28.5 & 7.7 & 1.312(17) & 1.258(17) & 0.18 & $-0.22(1)$ & $-0.21(3)$ & 2.5 & \cite{fkl+05,fsk+14}\\ 

J1906+0746 & NS & 144 & 3.98 & 1.323(11) & 1.290(11) & 0.08 &$-0.52(2)$ & $-0.565(6)$ & 5.4 & \cite{kasian}\\ 

B1913+16 & NS & 59.0 & 7.8 & 1.4398(2) & 1.3886(2) &  0.61 &$-2.402531(14)$ & $-2.396(5)$ & 9.9 &\cite{wnt10}\\ 

B2127+11C & NS & 30.5 & 8.0 & 1.358(10) & 1.354(10) & 0.18 &  $-3.95(13)$ &$-3.961(2)$ & 10.3(4) & \cite{jcj+06,kvf+14}\\ 

• & • & • & • & • & • & • & • & & &\\ 
 
J0348+0432 & WD &39.1 & 2.5 & 2.01(4) & 0.172(3) & $10^{-6}$ &$-0.258(11)$  & $-0.273(45)$  & 2.1(2) &\cite{afw+13}\\ 

J0751+1807 & WD & 3.4 & 6.3 & 1.26(14) & 0.13(2) & $10^{-6}$ & - & $-0.031(14)$ & 2.0 & \cite{lzc95,nsk08}\\ 

J1012+5307 & WD & 5.2 & 14.4 & 1.64(22) & 0.16(2) & $10^{-6}$&$-0.11(2)$ & $-0.15(15)$ & 0.836(80)& \cite{lwj+09,cgk98}\\ 

J1141$-$6545 & WD & 394 & 4.8 & 1.27(1) & 1.02(1) & 0.17 &  $-0.403(25)$ & $-0.401(25)$ &3.7 & \cite{bbv08,obv02a,abw+11}\\ 

J1738+0333 & WD & 5.9 & 8.5 & 1.46(6) & 0.181(7)  & $10^{-7}$ & $-0.028(2)$ &$-0.0259(32)$ & 1.47(10)& \cite{avk+12,fwe+12} \\ 

 & &  & • & • & • & • & • & & &\\

WD\,J0651+2844 & WD & - & 0.212 & 0.26(4) & 0.50(4) & 0 & $-8.2(17)$ & $-9.8(28)$& 1.0 & \cite{hkb+12a,bkh+11}\\

\end{tabular}

\end{center}
\end{minipage} }
\caption{Compact binaries in the Galaxy with measured intrinsic orbital decays.}
\end{table}

\section{The Hulse-Taylor Binary and the Double Pulsar}
The orbital decay due to emission of GWs was first detected in PSR\,B1913+16, the ``original'' binary pulsar discovered by Russell Hulse and Joe Taylor in the summer of 1974. The pulsar orbits an unseen NS every 7.8 hours with an eccentricity of $e = 0.62$. Soon after the discovery, the system's periastron  was found to change by 4.2\,deg\,$yr^{-1}$. Assuming the validity of GR, this change corresponds to a total mass of  2.83\,M$_{\odot}$. Together with $\gamma$, also measured soon after, the masses of the pulsar and the companion are constrained to $\sim 1.44$ and 1.39\,M$_{\odot}$ respectively. The detection of the orbital period decay was announced a few years later by Taylor {\it et al.} \cite{tfm79}. The measured value was in agreement with Einstein's quadrupole formula, providing direct experimental evidence for the existence of GWs. Perhaps not surprisingly, this result earned Hulse and Taylor a ticket to Stockholm in 1993. 

The most recent value of $\dot{P}_{\rm b}$ for the system agrees with GR at the $0.2\%$ level. While the experimental accuracy continues to increase with time, the precision of this radiative test is not expected to improve significantly. The main reason is the unknown distance to the binary and the uncertain Galactic gravitational potential at that location which limit the ability to correct the observed $\dot{P}_{\rm b}$ for the apparent kinematic contributions (see eq.\,7). 

Today we know of a more ``extreme'' system, that outshines the Hulse-Taylor binary in all aspects. 
PSR\,J0737$-$3039A/B, also known as the double pulsar, was discovered in a survey of the southern sky with the Parkes telescope in 2003 \cite{bdp+03,lbk+04}. The system has an orbital period of 2.45\,hours and an eccentricity of $e=0.088$, viewed at a remarkably edge-on inclination of $i=88.7$ degrees. Furthermore, both NSs are seen as pulsars allowing to infer the theory-independent mass ratio from the observed Roemer delays. 
In total six pK parameters are measured; five in the timing data of pulsar A  \cite{ksm+06} and an additional  from the effects of geodesic precession on pulsar B \cite{bkk+08}. 
These quantities, together with the mass-ratio allow for $7-2 = 5$ independent tests of gravity, which all  agree with GR. 

Because of its proximity to Earth, its distance is directly measured with Very-Long Baseline interferometry \cite{dbt09}. This allows to correct the observed $\dot{P}_{\rm b}$ for kinematic contributions with much better precision compared to the Hulse-Taylor binary. The latest published value \cite{ksm+06} agrees with GR at the 1.4\% level, while the current  value (to be published soon by Kramer {\it et al.}) reaches a precision below 0.05\%. 
The double pulsar provides by far the best verification of Einstein's quadrupole formula. Furthermore, the high-precision of the experiment rules out a range of alternatives such as Beckenstein's TeVeS (see Section\,9).

\begin{figure}[h]
  \begin{center}
   \label{fig:2}
    \includegraphics[width=0.9\textwidth]{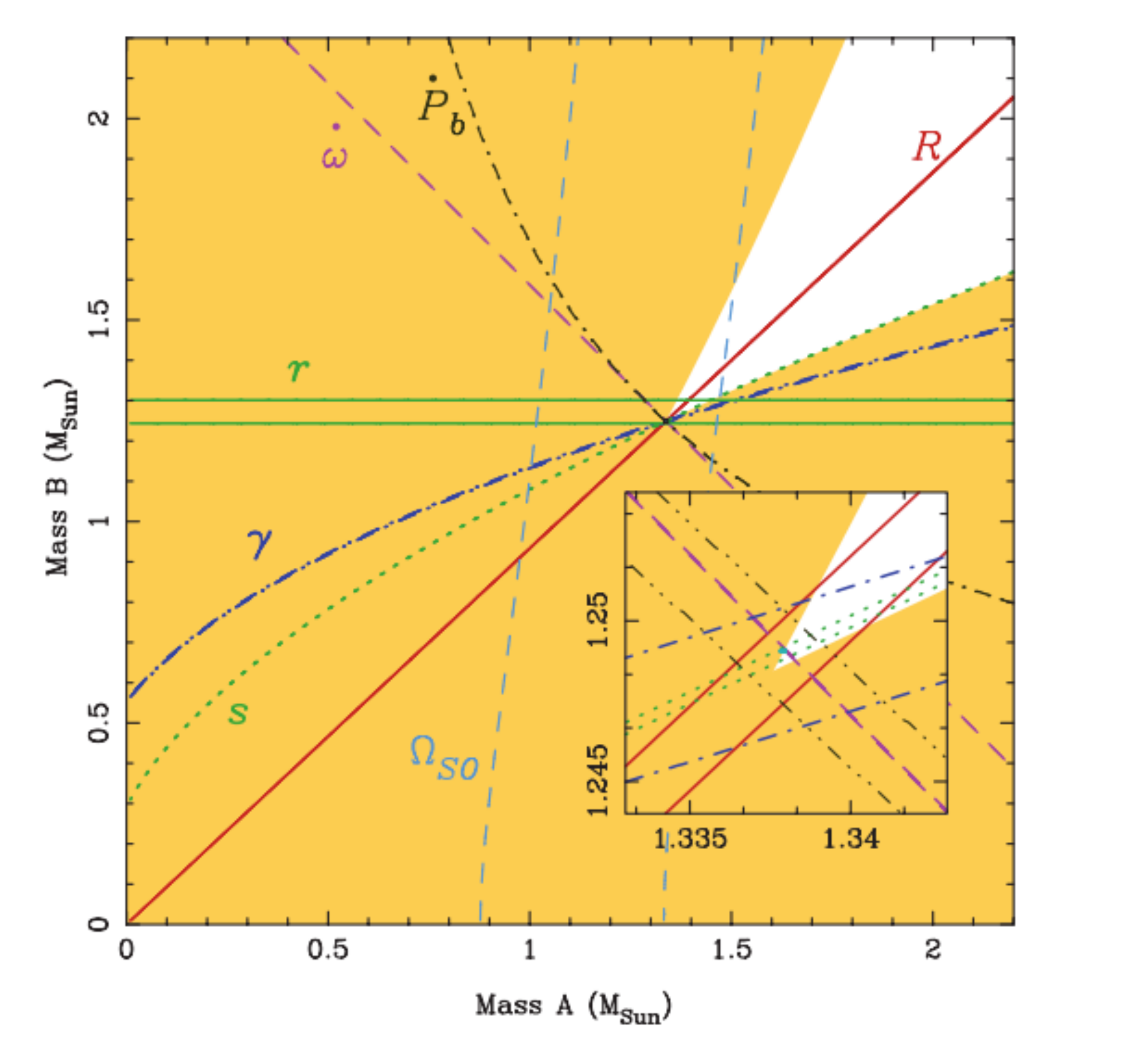}
     \caption{Post-Keplerian effects and corresponding $1\sigma$ uncertainties for the double pulsar system PSR\,J0737$-$3039 as of 2011 (courtesy of Michael Kramer).}
  \end{center}
\end{figure}

\section{Beyond the Double Pulsar} 

The orbital decay of J0737$-$3039 leaves very little room for deviations from  Einstein's quadruple formula. 
Unfortunately, this doesn't necessarily guarantee the absolute validity of GR for the energy scales probed by orbiting NSs. Many alternatives to GR 
\emph{also} predict the same rate of GW emission, even if NSs behave very differently. In many (metric) alternatives, the gravitational interaction is mediated not only by a tensor field  but also by additional fields, for example scalar or vector.  These fields usually break local symmetries leading to a SEP violation and/or a violation of the Local Lorenz Invariance for the gravity sector. Such effects  lead to a number of modifications to the binary orbit. For the class of compact binaries discussed here the most prominent effect is the emission of \emph{dipolar} gravitational radiation, on top of the quadrupole and higher-order terms, also present in GR. The associated orbital decay is a 
$\mathscr{O}(v^3/c^3)$ post-Newtonian effect that reads \cite{de96a,wex}:
\begin{equation}
\dot{P}_{\rm b}^{\rm dipole} = - \frac{4 \pi ^2 G_{\ast}M_{\odot}}{c^3 P_{\rm b}}\frac{m_{\rm p}m_{\rm c}}{m_{\rm p} + m_{\rm c}}  \frac{(1 +\, {e^2}/{2} )}{ (1
-e^2)^{5/2}}   (\alpha_{\rm p} - \alpha_{\rm c})^2
\end{equation}
where $G_{\ast}$ is the bare gravitational constant and  other symbols have their usual meaning. The terms $\alpha_{\rm p}$ and $\alpha_{\rm c}$ are the effective coupling terms that depend on the coupling strength of the extra fields to matter inside the stars and usually scale non-linearly with the binding energy. Obviously in GR, $\alpha_{\rm p} = \alpha_{\rm c} = 0$, but in some alternative theories these terms can reach unity \cite{de93,gw02,esp04}. 
Similarly to the quadrupole term, dipole radiation depends sensitively on the orbital period. A fundamental difference however is that the associated orbital decay depends on  $\alpha_{\rm p} - \alpha_{\rm c}$ which might take completely different values for the two stars. It should now be obvious why double NSs make poor laboratories for testing the existence of this effect. In these systems the compositions of the two stars are almost identical, meaning that  $(\alpha_{\rm p} - \alpha_{\rm c})^2\simeq 0$, even if $\alpha_{\rm p}$ is significantly larger than zero. Therefore, dipole radiation is not relevant for DNSs, even if there is an underlying modification to gravity that forces stars to respond differently to gravity compared to GR. 

Evidently, the best laboratories to test the predictions of these theories are tight binaries where the companion star is expected to have $\alpha_{\rm p} \neq \alpha_{\rm c}$. Such 
asymmetric systems are in fact not rare. The advent of recent radio surveys has reveled a rich ensemble of binary pulsars with main-sequence, semi-degenerate and WD
companions. Most of these systems however either have too long orbital separations or their orbital motion is severely influenced by classical phenomena, such as movement of material due to 
irradiation, tides and quadrupole-moment variations \cite{lvt+11}. A fortunate exception to this picture are pulsars with WD companions. WDs have extremely small gravitational binding energies 
compared to NSs, but yet they are ``compact'' enough to be treated as point masses. Presently we know five pulsar--WD binaries with orbital periods less than a day. 
These systems are the end-point of evolution of stars that survived a common envelope phase, followed by a long period of stable mass transfer. As we shall see bellow their formation -- 
 not yet fully understood -- gives them a set of salient orbital and intrinsic  properties that make them exquisite labs for strong-field gravity. We should also note that the 
\emph{ideal} system to test dipolar GW emission would be a pulsar orbited by a black hole. Unfortunately, such a system is yet to be found. 

\section{Low-Mass White Dwarfs in Binaries} 
Low-mass white dwarfs (LMWDs)  represent the final evolutionary product of  stars with masses $\le 0.4$\,M$_{\odot}$ that cannot evolve beyond the helium flash. Single stars with 
such low mass evolve extremely slowly and have in fact not yet settled on the WD cooling track. However, formation of LMWDs is still possible in binaries where higher-mass stars can 
loose significant amounts of material  throughout their life. LMWDs are the most common counterparts to millisecond pulsars, a fact that strongly supports that the high-angular 
momentum of these NSs is a consequence of mass and angular momentum transfer from the LMWD progenitors. These binaries possess numerous interesting properties, studied extensively over the past few decades (e.g. \cite{dsbh98,ts99a,sar+02,pac07}). Those most relevant to gravity tests are the following: 

\begin{enumerate}
\item{{\bf LMWDs are not exactly WDs.}} Textbook WDs are objects where all nuclear-fusion reactions have ceased and the entire stellar luminosity is provided by the latent heat of the core. For most LMWDs, this is not the case. When these stars settle on their cooling tracks they may still have an extensive amount of hydrogen surrounding their 
degenerate helium cores \cite{dsbh98}. In most cases the pressure provided by the envelope leads to conditions that allow residual hydrogen burning close to the degenerate core. 
Nuclear burning becomes the dominant energy and allows the stars to evolve on nuclear rather than thermal timescales, which is the case for higher-mass WDs. For gravity tests this is 
important because it allows to study some of these stars spectroscopically and infer their masses. 
\item{\bf LMWDs flash.} The initial envelope mass  depends mainly on the orbital separation during detachment, i.e. it is a strong function of the stellar mass. The less massive the WD 
the larger the (relative) size of the envelope. Therefore, because the size of the nuclear burning region also depends on the envelope size, there exists a range of masses for which nuclear burning is unstable. This thermal instability develops because the shell is not thick enough to compensate sudden changes in temperature by regulating its density, like normal 
stars do. For a shell of thickness $d = r - r_{\rm 0} \ll R$, fractional changes in temperature and density are related via:
\begin{equation}
\frac{\delta T}{T} = \frac{1}{\chi_{T}}\left( 4 {\frac{d}{r} - \chi_{\rho}}\right) \frac{\delta \rho}{\rho}
\end{equation}
where $\chi_{T} > 0$ and $\chi_{\rho}$ are the logarithmic pressure gradients under constant temperature and density respectively \cite{kw90}. From Eq.\,10 it follows that a shell is thermally 
unstable when $d < r\chi_{\rho}/4$. In LMWDs the most severe flashes are those induced by unstable CNO burning (rate $\propto T^{18}$) shortly before the star enters the cooling 
track. CNO flashes may consume large amounts of hydrogen resulting to complete cessation of nuclear reactions afterwards. Furthermore, during a flash episode, the radius of the star 
may expand beyond its Roche lobe causing further decrease of its mass. Modern stellar evolution calculations place the lower mass limit for CNO-induced flashes  between $0.17$ and 
$0.21$\,M$_{\odot}$, with  the exact value depending on the envelope metallicity, rate of gravitational settling and treatment of microphysics. Unfortunately knowledge of the 
critical mass value is crucial for determining LMWD masses and is one of the main limiting factors for gravity tests. Recent discoveries of high-temperature LMWDs  with masses $< 0.2$\,M$_{\odot}$ strongly suggest that the threshold for flashes is indeed above this value (see e.g. \cite{bka+10}.  This  is further supported by observations of the system 
PSR\,J1909$-$3744 for which the mass and radius are measured independently \cite{antoniadis}. Hence,
assuming that the 0.2\,M$_{\odot}$ threshold for flashes is universal, then all WDs with lower masses should still possess thick hydrogen envelopes.
 
\item {\bf Pulsars with LMWD companions have extremely circular orbits.} A final remark is that tidal dissipation during the mass-transfer phase circularizes the orbit on very short timescales. For gravity tests this becomes an issue because it makes the two PK parameters that yield precise masses
for compact, eccentric binary pulsars ($\dot{\omega}$ and $\gamma$,
eqs. 2 and 5) unmeasurable. Thus, masses can be inferred from timing only for orbits sufficiently edge on, which is not the case for any of the known compact binaries. Consequently alternative mass-determination methods should be applied, some of which are further discussed in the following Section.
\end{enumerate}

\section{White Dwarf Masses}
Strong-field gravity tests with binary pulsars require  knowledge of the stellar masses.  For compact binaries with small eccentricities the 
only way to achieve this through timing is by measuring the Shapiro-delay signature of the pulsar signal as it passes through the (weak) gravitational potential of the companion. Since, 
this requires  sufficiently edge-on viewing, it can only be applied to very limited cases. In fact, none of the known relativistic binaries has a sufficiently edge-on inclination to allow for a 
clear measurement of the Shapiro delay. However, because LMWDs stay hot and bright for a significant fraction of their lifetimes, one can use  methods based on optical observations to 
infer their masses. These methods have the advantage that the derived masses are independent of any specific gravity theory.

The most straightforward technique involves phase-resolved spectroscopy of the WD. Doppler measurements of the WD's Balmer lines along the orbit, trace the orbital motion, the 
amplitude of which is directly proportional to the size of the WD orbit. Because the size of the pulsar's orbit is also measured via radio timing, the ratio of the two quantities gives the 
mass ratio, $q$ of the system. Furthermore, one can also compare the observed spectrum with atmospheric models. Atmospheric Balmer lines have shapes and depths that can be 
modeled using only two parameters, the effective temperature $T_{\rm eff}$ and surface gravity $g = Gm_{\rm c}/R_{\rm WD}^2$ \cite{spectra}. Once these two quantities are 
determined  the WD 
mass and radius can be inferred using an appropriate finite-temperature mass-radius relation (see Section\,6). An additional side-product is the measurement of the systemic 
radial velocity which together with the proper motion of the system and the distance, make possible to infer the 3D motion of the binary in the Galaxy. 
A drawback, apart from the dependence on theoretical mass-radius relations, is the usage of model atmospheres. Although LMWDs are extremely simple, atmospheric models based on 
1D calculations are known to yield biased values for the surface gravity if the outer envelope is convective \cite{antoniadis}. However,  errors can be accounted for with appropriate 3D 
models \cite{tls+13} and vanish completely for hot, radiative atmospheres.  

Another method that does not depend on atmospheric models relies on the parallax measurement of the system \cite{avk+12}. Once the distance is known with sufficient precision, one can infer the absolute stellar luminosity. Together with the temperature this yields the WD radius, which again combined with a mass-radius relation, give the WD mass. This method has not yet yield any precise mass measurements, but it may be proven useful in the near future. 

The first compact binary studied with spectroscopy was PSR\,J1012+5307, a 5.2\,ms pulsar in a 14.4 hours orbit. Because of the large uncertainties in atmospheric models at that time, the mass was pinned down with a precision of only $\sim 12\%$ \cite{vbk96,cgk98}. Nevertheless, together with a $\sim 1\%$ detection of the orbital period decay, which is agreement with the prediction of GR \cite{lwj+09}, the system places a constraint on dipole radiation of $\dot{P}_{\rm b}^{\rm dipole} = (-0.2 \pm 1.6)\times 10^{-14}$. 
Presently the system is superseded by two similar systems, discussed in the remaining of this chapter.

\section{Constraints on Dipolar Gravitational Radiation}
\subsection{PSR\,J1738+0333, the most stringent constraints on dipolar radiation}
PSR\,J1738+0333 is a 5.9\,ms pulsar in a 8.5\,hours circular orbit around a LMWD. It was discovered in 2001 with the Parkes telescope in Australia and is being timed ever since with Arecibo, Effelsberg and Westerborg telescopes, on a monthly basis. The long baseline of precise timing observations yields highly precise measurements of the spin, kinematic and orbital parameters, including the parallax and systemic proper motion \cite{fwe+12}. These quantities allow to compute the Shklovskii contribution to the orbital decay, $\dot{P}_{\rm b}^{\rm Shk} = 8.3^{+0.6}_{-0.5}$\,fs\,s$^{-1}$. Together with the measured orbital decay $\dot{P}_{\rm b} = -17.0\pm 3.3$\,fs\,s$^{-1}$, this implies that the intrinsic orbital period decay is $\dot{P}^{\rm int}_{\rm b} = -25.9 \pm 3.2$\,fs\,s$^{-1}$. 

Optical observations yield a mass ratio of $q = 8.1\pm 0.2$, a WD mass of $ 0.181^{+0.007}_{-0.005}$\,M$_{\odot}$ and a pulsar mass of  $1.46$\,M$_{\odot}$ \cite{avk+12}. The latter are based on the assumption that the WD envelope is thick, which is most likely the case, given the high temperature  and the slightly lower mass compared to the
WD in PSR\,J1909$-$3744. Given the masses, the rate of orbital decay according to GR (Eq.\,6) should be $-27.7^{+1.5}_{-1.9}$\,fs\,s$^{-1}$ 
which is in agreement with the observed value. Since both  stars can be treated as point particles \cite{fwe+12}, the ``excess'' orbital decay of $+2.0^{+3.7}_{-3.6}$,fs\,s$^{-1}$ 
directly translates to a constraint on deviations from the quadruple formula. Apart from dipole radiation, these may already enter at the Newtonian level, due to a varying gravitational constant $\dot{G}$. Lunar Laser Ranging experiments yield a limit of $\dot{G}/G$=$(−0.7 \pm 3.8) \times
10^{-13}$\,yr$^{-1}$, thus conservatively one can assume: $\dot{P}_{\rm b}^{\dot{G}} = -2P_{\rm b}\dot{G}/G = (+0.14 \pm 0.74)$\,fs\,s$^{-1}$. Therefore, the limit on dipole radiation from this test is $\dot{P}^{\rm dipole}_{\rm b} = 1.9^{+3.8}_{-3.7}$\,fs\,s$^{-1}$, the most stringent obtained so far. 
Since GR predicts  $\dot{P}_{\rm b}^{\rm dipole}=0$, the theory passes the test. As we shall see below, this is not the case for many alternative theories.

\subsection{PSR\,J0348+0432, a massive pulsar in a compact binary}
This extraordinary system was discovered in a drift-scan survey \cite{drift,discovery_paper} conducted with the Green-bank Telescope in West Virginia. The pulsar appears to be "mildly recycled", with a spin period of 39\,ms and relatively strong surface magnetic field of $B=3\times10^9$\,Gauss. The NS orbits a LMWD companion every 2.46\,hours. The orbital period --- by far the shortest among this type of systems --- is only 15 seconds longer than that of the double pulsar. Consequently, the orbital period decays relatively fast and it is easier to measure precisely on short timescales. 

Spectroscopic observations of the WD companion \cite{afw+13} were conducted with the FORS2 instrument of the VLT in Chile. The data show an effective temperature of $\sim 10100$\,K and a surface gravity of $\log g  \sim 6.0$\,dex. Independently of any model, the high temperature constraint the WD mass to a narrow interval: the star cannot be less massive than $\sim 0.165$\,M$_{\odot}$ as WDs with lower masses never reach such high temperatures on the cooling track. Furthermore, if its hydrogen envelope were thin, it would have cooled to this temperature in only few Myrs, which would imply that there is a very high abundance of these stars, given the 13\,Gyrs age of the Universe. 
Using appropriate cooling models with thick envelopes the WD mass is constrained to $0.172\pm0.003$\,M$_{\odot}$. Combined with the measured mass ratio, $q=11.70\pm0.13$, the inferred pulsar mass is $2.01\pm0.04$\,M$_{\odot}$, the highest NS mass measured until the time of writing. 

High NS masses are important for constraining the EoS of supra-nuclear matter, as many models place the dividing line between NSs and black holes at lower values. A two-solar mass NS is also important for gravity tests as it has a gravitational binding energy $\sim 50\%$ higher than lower-mass NS like PSR\,J1738+0333 (in GR). As the magnitude of deviations from GR can depend non-linearly on the binding energy, these stars can be used to probe a strong-field gravity regime which is qualitatively very different compared to other binary-pulsar experiments (Fig.\,3).

Since April 2011 the system is being monitored monthly with Arecibo and Effelsberg.  To match the arrival times, the orbital solution requires a significant rate of orbital decay, $P_{\rm b}^{\rm obs} = (-273\pm45)$\,fs\,s$^{-1}$. The former is largely unaffected by kinematic contributions which are an order of magnitude bellow the measurement error. The GR prediction given the masses is $-258^{+8}_{-11}$\,fs\,s$^{-1}$, in agreement with the observed rate. The $20\%$ precision of this GR test is quantitatively poor compared to the remarkable precision achieved for e.g. the double pulsar. However, because of the high pulsar mass, the resulting constraints on some alternative theories are infinitely more stringent.
\begin{figure*}[h]
  \begin{center}
   \label{fig:3}
    \includegraphics[width=0.85\textwidth]{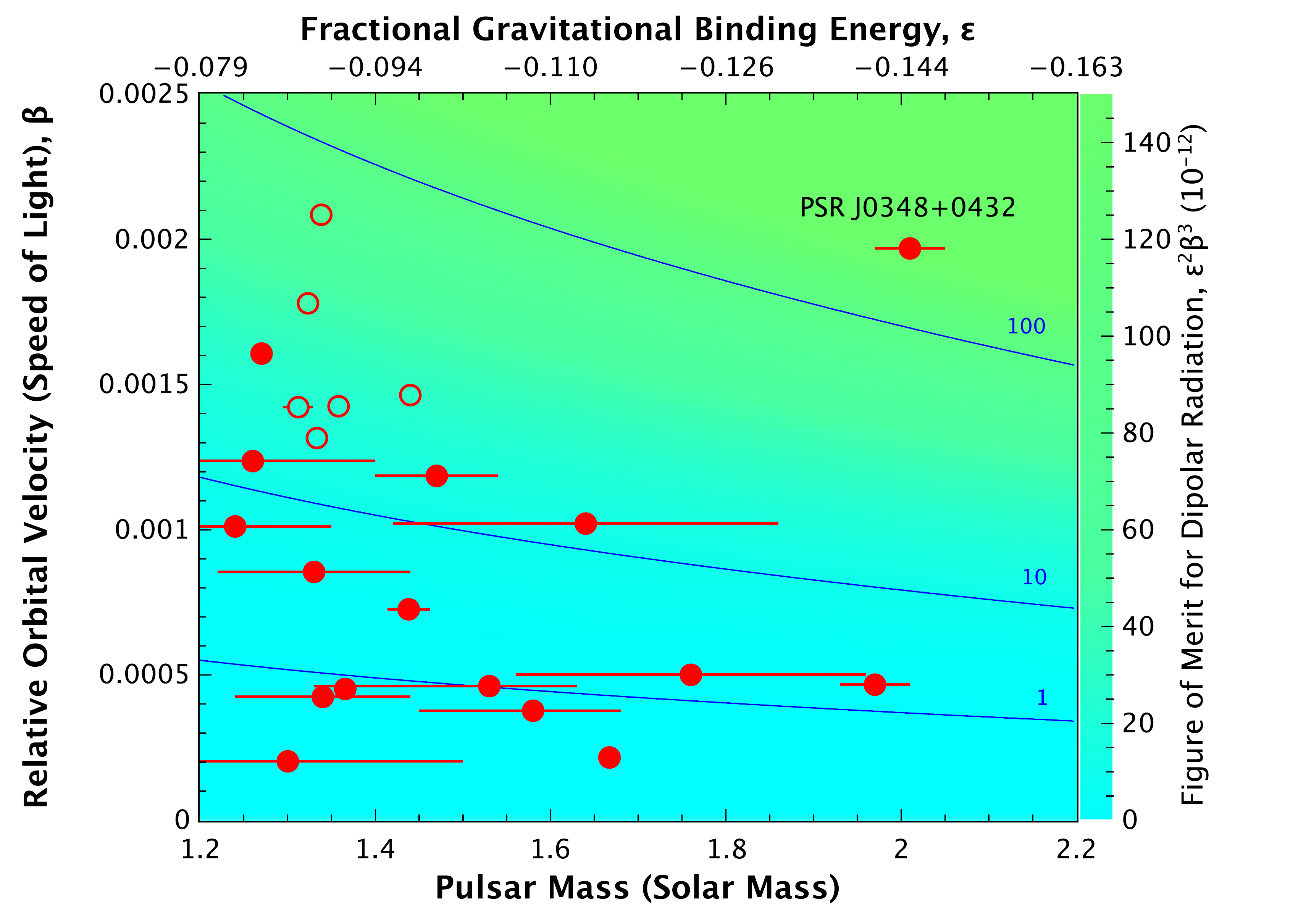}
     \caption{Mean orbital velocity as a function of NS mass for the compact binary pulsars of Table\,2. 
     Hollow circles represent double NS systems where no significant dipolar emission is expected. PSR\,J0348+0432 stands out as the most ``luminous'' emitter. Courtesy of Norbert Wex.}
  \end{center}
\end{figure*}

\section{Ramifications for the Phase Evolution of Neutron Star Mergers}  
In-spiralling compact binaries will be the main source of GW detectable by the forthcoming ground-based detectors. The  GW signal from merger events will be deeply buried in the detectors' noise, requiring matched filtering techniques for extraction \cite{ss09}.  Ideally, theoretical templates used for detection should track the phase evolution of the merger with a precision better than 1 cycle over the $\sim10^4$ orbits covering the detectors' band. Presently, models calculated by various groups have reached a precision of  $\mathscr{O}(v/c)^7$, with the current forefront of research being pushing towards the 4th post-Newtonian order \cite{djs14,bla14}. 
\begin{figure}[h]
  \begin{center}
   \label{fig:4}
    \includegraphics[width=0.85\textwidth]{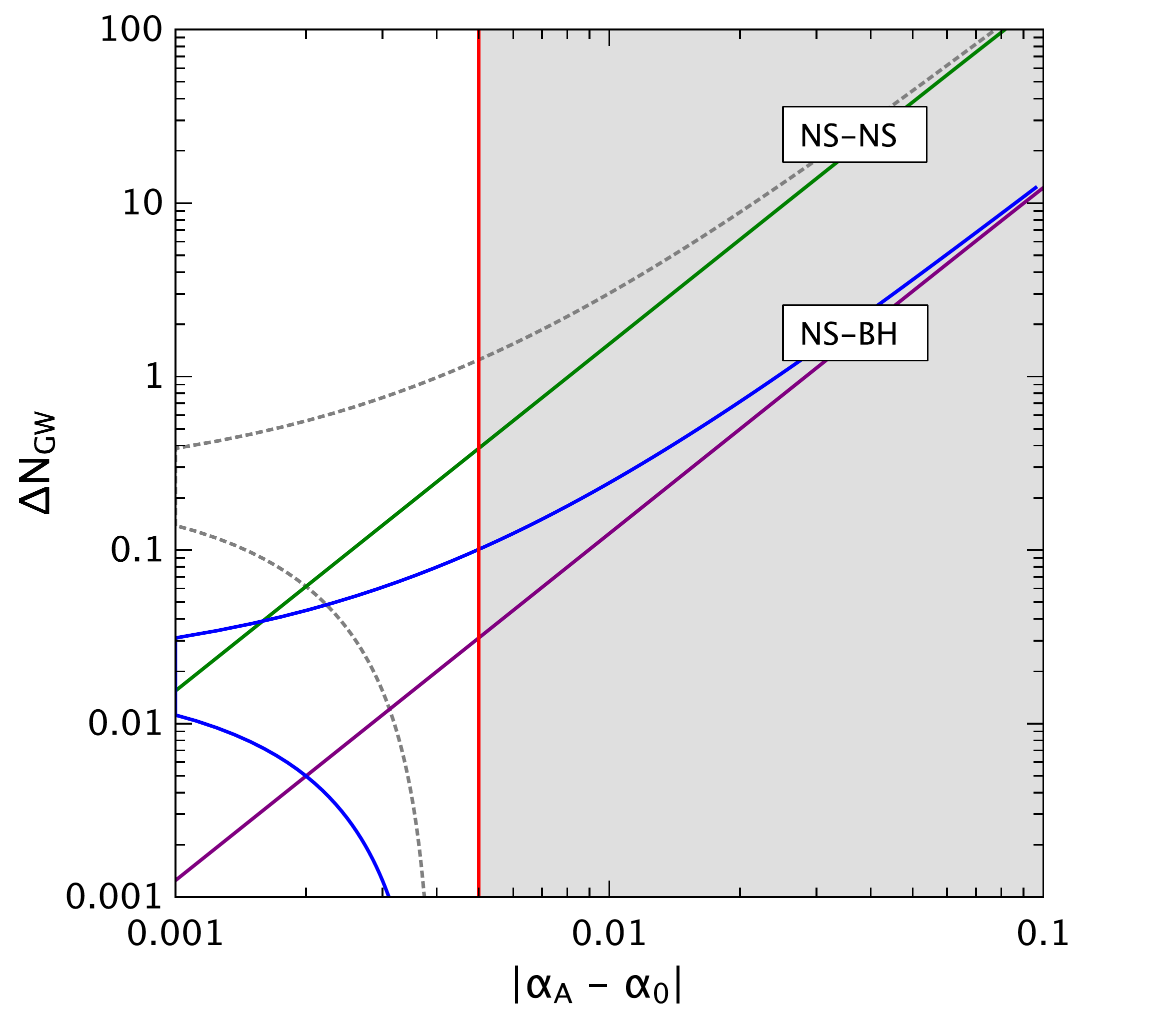}
     \caption{Maximum phase offset in the band 20\,Hz to few kHz between the GR template and the true in-spiral in the presence of dipolar radiation, as a function of the scalar coupling strength. The two examples show the merger of a 2\,M$_{\odot}$ NS with a 1.25\,M$_{\odot}$ NS (green) and a 2\,M$_{\odot}$ NS with a 10\,M$_{\odot}$ black hole. The gray area is excluded by the limits from binary pulsars (for details see \cite{afw+13}).}
  \end{center}
\end{figure}
However if the properties of gravity differ from the GR prediction, loss of energy to dipole GWs may drive the merger many cycles away from the templates, jeopardizing detection. This is true even for double NSs that have small differences on their binding energies.  
The limits on dipolar radiation imposed by PSRs\,J1738+0333 and J0348+0432 imply $|\alpha_{\rm p} - \alpha_{0}| \le 0.005$ for the whole range of NSs observed in nature. 
Consequently, mergers cannot loose more than half a cycle compared to GR,  supporting the use of GR templates. We should note that this may not be true if the partner particle to graviton is massive, although PSR\,J0348+0432 already constraints its mass above $\sim 10^{-13}$\,$e$V/$c^2$.

\section{Constraints on Alternative Theories of Gravity}

\subsection{Scalar-Tensor Gravity}
Scalar-Tensor theories of gravity are  the most natural extensions of GR, where the gravitational interaction involves an additional scalar-field contribution $\phi$, in addition to the spin-2 graviton. 
In these theories, matter is assumed to be coupled to a physical metric $\hat{g}_{\mu \nu} \equiv A^2(\phi)g_{\mu \nu}^{\ast}$, where $A(\phi)$ is an generic function of the matter-scalar 
coupling and $g_{\mu \nu}^{\ast}$ is the  pure spin-2 field (Einstein metric). Since $A(\phi)$ is a-priori unknown, it is useful to expand it around the cosmologically imposed scalar field $\phi_{0}$ as $\ln A(\phi) =  \alpha_{0}(\phi - 
\phi_{0}) + 1/2\beta_{0}(\phi - \phi_{0})^2 + \ldots$ \cite{de92,de93,de96,de96a,de98,esp99,esp04}. Here, $\alpha_{0}$ and $\beta_{0}$ denote the linear and quadratic coupling terms respectively and higher-order terms are 
neglected. This notation allows to embed GR and other well-studied theories in the space of scalar-tensor alternatives. For example in GR, $\alpha_{0} = \beta_{0} = 0$  and in the 
Jordan, Fierz, Brans-Dicke theory, $a^2_{0} = 1/(2\omega_{\rm BD} + 3), \beta_{0}=0$ \cite{wex}. In the strong-gravity potential of a NS, $\alpha_{0}$ and $\beta_{0}$ are modified by strong-gravity effects and become body-dependent  ``charges'' $\alpha_{\rm p,c}$ which can be computed  by means of numerical integration of the modified hydrostatic equation, assuming an EoS. For binary pulsars with WD companions $\alpha_{\rm c} \simeq \alpha_{0}$ which is constrained below $\sim 4\times 10^{-4}$ by Solar-System experiments. Therefore, limits on dipolar radiation can be viewed as direct constraints on $\alpha_{\rm p}$ that can reach unity for specific (negative) values of $beta_{0}$ due to non-perturbative effects \cite{esp04}.

\begin{figure*}
$\begin{array}{cc}
\resizebox{5.7cm}{!}{\includegraphics{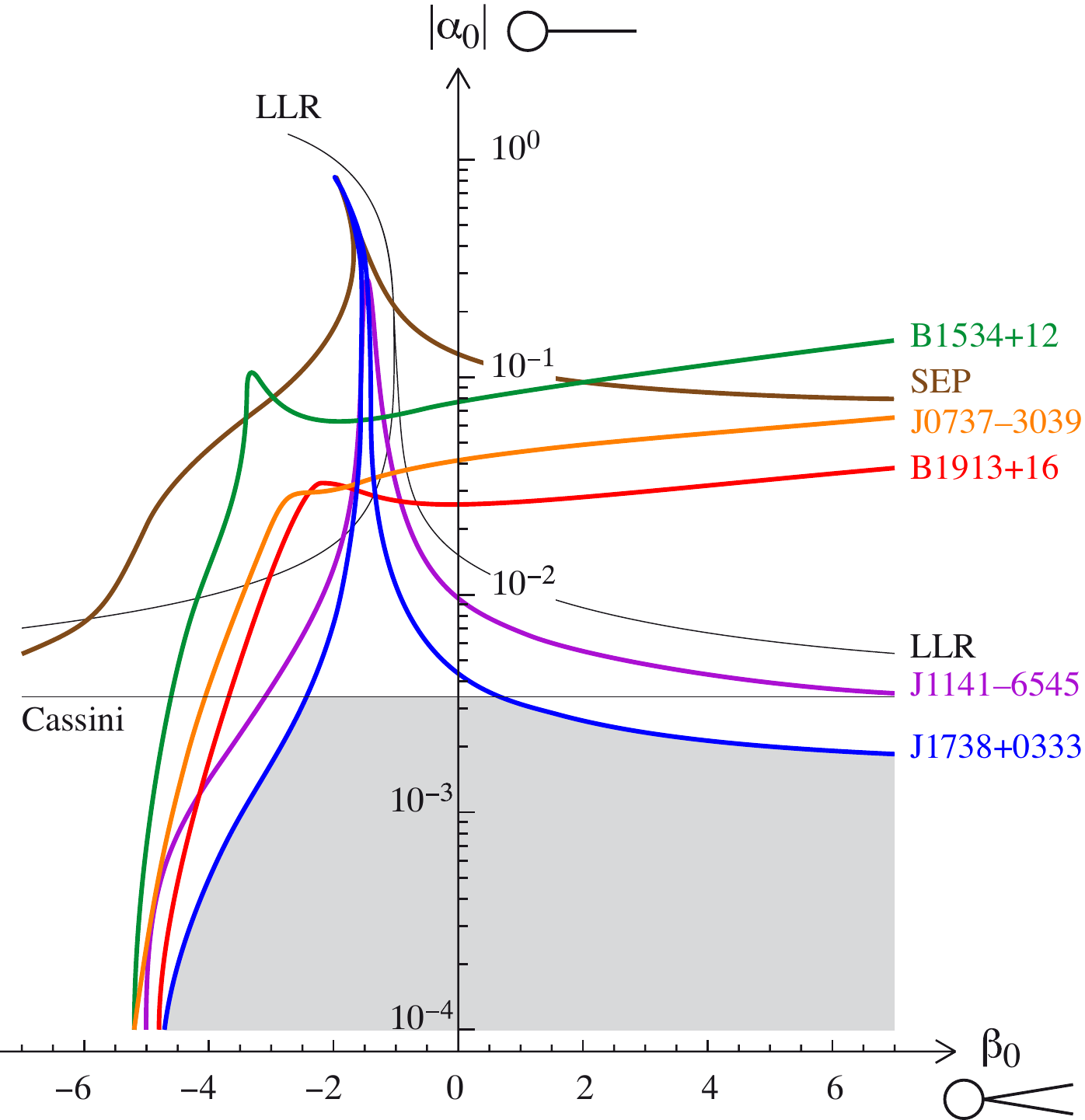}} &
\resizebox{5.7cm}{!}{\includegraphics{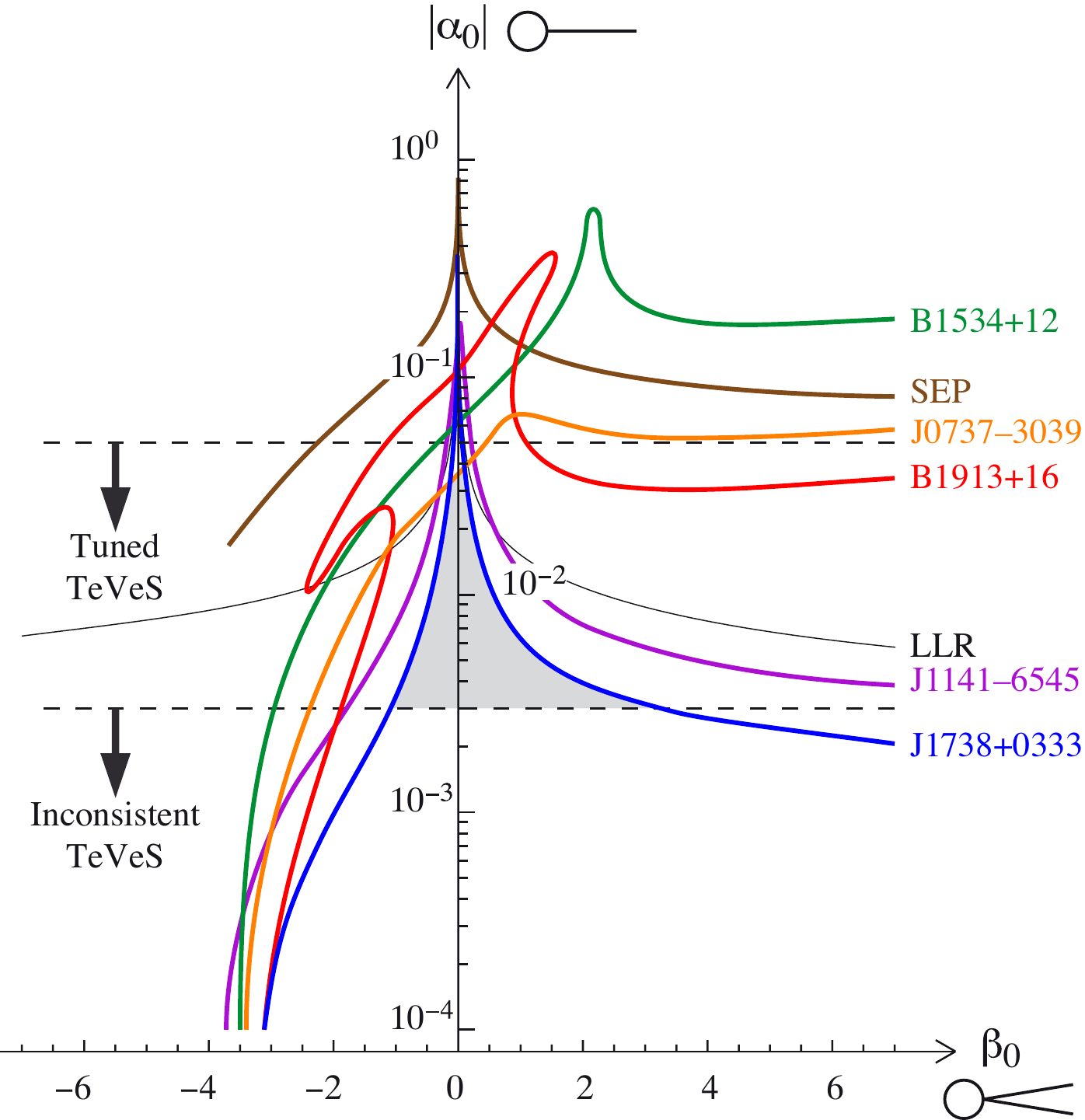}}
\end{array} $
\caption{ \textit{Left:} Constraints on the linear and quadrupole scalar-matter coupling in Scalar-Tensor Gravity. Coloured lines show various binary pulsar and Solar-System experiments. The shaded area is the regime still consistent with all experiments. \textit{Right:} Constraints on scalar-matter coupling in the TeVeS framework. \cite{fwe+12}}
\label{fig:5}
\end{figure*}
Figure\,5 shows the limits on $\alpha_{0},\beta_{0}$ imposed by binary pulsar experiments and various Solar-System tests. For most of the parameter space, PSR\,J1738+0333 poses the most significant constraints. This is especially true for negative values of $\beta_{0}$, where non-perturbative effects are more prominent. For the original Jordan-Brance-Dicke scalar-tensor theory the Cassini bound is still more constraining by a factor of $\sim 2$, but this may change as more data for PSR\,J1738+0333 become available.  

Because of its high mass, PSR\,J0348+0432  provides a complementary constraint for negative values of $\beta_{0}$ still consistent with PSR\,J1738+0333. Due to the high gravitational binging energy inside the NS, $\alpha_{p}$ may assume order-unity values even for $\alpha_{0}=0$. The timing data impose a limit of $|\alpha_{p} - \alpha_{\rm c}|<5\times 10^{-3}$, therefore excluding ``spontaneous scalarization'' even for 2\,M$_{\odot}$ NSs.  

Similar constraints are obtained for certain $f({ R})$ theories that show very similar behaviour \cite{dl13,dl13b}. 

\subsection{TeVeS}
Tensor-Vector-Scalar (TeVeS) gravity is a relativistic formulation of Modified Newtonian Dynamics (MOND) that seeks to explain galaxy rotation curves and week lensing without invoking Dark Matter \cite{mm83}. 
In MOND this is achieved by modifying Newton's equation below a fundamental acceleration scale $a_{0} \simeq 10^{-10}$\,m\,s$^{-2}$. In TeVeS  gravity is mediated by tensor, vector and scalar fields. The scalar-field kinetic term is an unknown non-linear function, which has a natural shape only if $|\alpha_{0}| > 0.05$, 
otherwise the theory would be inconsistent with Solar-System constraints \cite{fwe+12}. Below this value the function becomes highly fine-tuned and completely unnatural (bi-valued) for 
$|\alpha_{\rm 0} |<0.003$. If one assumes a generalized scalar-matter coupling just like in scalar-tensor theories, binary pulsar experiments impose the limits shown in the right panel 
of Figure\,5 \cite{fwe+12}. For most of the parameter space, the theory predicts significant amounts of dipolar radiation and therefore PSR\,J1738+0333 posses the most stringent 
constraints. For the original theory however where $A(\phi) = \exp(\alpha_{0}\phi)$  there is no low-order modification to the gravitational radiation damping. However, modifications still enter the orbital dynamics through other pK effects, thus the much more precise (published) double-pulsar data pose the most stringent constraints. As it can be seen in Fig.\,5, TeVeS is already confined within the ``unnatural'' regime. 

\subsection{Local Lorenz Invariance}
Local Lorenz Invariance violations (LLIV) in the gravity sector would introduce a preferred frame at each spacetime point, that would alter the orbital dynamics of binary pulsars.
In the parametrized post-Newtonian framework LLIV is characterized by two parameters $\alpha_{1,2}$ and an additional $\xi$ (Whitehead) parameter that quantifies the local position invariance \cite{will06}.  In the general case where the binary moves relative to the preferred frame with 
velocity ${\bf w}$, $\alpha_{1}$ would induce an orbital polarization and $\alpha_{2}$ a change of the projected semi-major axis due to the precession of the orbital plane \cite{sw12}. 
Assuming that the preferred frame coincides with the Cosmic-Microwave background, PSRs\,J1738+0333 and J1012+5307 placed a stringent limit of $\hat{\alpha}_{1} = 0.4^{+3.7}_{-3.1} 
\times 10^{-5}$ \cite{sw12}, which represents a significant improvement over Solar-System tests. Here, the ``hat'' represents possible modifications due to strong-gravity effects. The same binaries limit $|\hat{\alpha_{2}}|$ below $1.8\times10^{-4}$ \cite{sw12}. This 
parameter is in fact better constrained from studies of the profile stability in solitary pulsars ($|\hat{\alpha_{2}}| < 1.6\times10^{-9}$, \cite{sck+13}), which also provide the most stringent limit on $|\hat{\xi}|$ ($< 3.9 \times 10^{-9}$) \cite{sw13g}.
 LLIV would also result to emission of dipolar waves. Several studies have used the limits from the aforementioned binaries to place stringent constraints on LLIV theories such as Einstein-\AE ther and khronometric gravity \cite{ybyb14,ybby14}.

\section{Final Remarks} 
Present timing experiments have reached remarkable sensitivities and place stringent constraints on a wide range of possible deviations from GR. 
Binary pulsars are often regarded as tests of gravity at the binary-separation scale. While this is certainly true for theories like GR, a wide range of possible strong-field deviations  in 
the non-linear  environment of the NS interior could propagate over large distances, making binary pulsars sensitive (and thus-far unique) laboratories for strong gravity. This is 
especially true when considering GW emission which was the focus of this chapter.  

In view of the stringent limits summarized here and the anticipated operation of  advanced ground-based detectors, it may be tempting to ask whether if it is necessary to continue searching for modified gravity in  pulsar experiments. Arguably, the answer is positive, both for philosophical and practical reasons. GR, is one of the  fundamental pillars of modern physics and therefore deserves to be tested in every way possible. 
Furthermore, tests of gravity have a deep theoretical motivation since GR is incompatible with quantum mechanics and therefore {\it should} break down beyond some energy scale. It also possesses  a number of pathological properties like for instance the onset of singularities in gravitational collapse. 
Since the magnitude and scale of possible deviations is a-priori unknown, it thus makes sense to both continue improving the sensitivity of current experiments and also seek additional systems able to provide complementary tests. For example, improving the precision of the double pulsar experiment  may provide tests of the orbital dynamics at higher post-Newtonian orders.
Similarly, systems like a pulsar-black hole binary or the newly discovered triple pulsar \cite{fkw12,rsa+14} may yield a many-fold sensitivity improvement in the search for SEP violations and novel tests of other fundamental properties like the no-hair theorem \cite{lwk+12}. Finding deviations from GR would be a major
advance in our understanding of the Universe and represent a vital pointer towards a more fundamental description of the laws of nature.

\begin{acknowledgement}
I gratefully acknowledge financial support by the European Research Council for the ERC Starting Grant BEACON under contract no. 279702 (PI: Paulo Freire). I also wish to thank Paulo Freire, Norbert Wex and Michael Kramer for  thorough discussions, insightful comments and ideas. This research has made extensive use of NASA's Astrophysics Data System (ADS).
\end{acknowledgement}

\bibliography{author}

\end{document}